\documentclass
[preprint,byrevtex,10pt,twocolumn,tightenlines,prl,letterpaper,noshowpacs]{revtex4}%
\usepackage{amsfonts}
\usepackage{amsmath}
\usepackage{amssymb}
\usepackage[dvips]{graphicx}
\usepackage{hyphenat}%
\providecommand{\U}[1]{\protect\rule{.1in}{.1in}}

\begin{document}
\preprint{ }
\title{Long-range influence of manipulating disordered-insulators locally}
\author{Z. Ovadyahu}
\affiliation{Racah Institute of Physics, The Hebrew University, Jerusalem 91904, Israel }


\begin{abstract}
Localization of wavefunctions is arguably the most familiar effect of disorder
in quantum systems. It has been recently argued [V. Khemani, R. Nandkishore,
and S. L. Sondhi, Nature Physics, \textbf{11}, 560 (2015)] that, contrary to
naive expectation, manipulation of a localized-site in the disordered medium
may produce a disturbance over a length-scale much larger than the
localization-length, $\xi$. Here we report on the observation of this nonlocal
phenomenon in electronic transport experiment. Being a wave property,
visibility of this effect hinges upon quantum-coherence, and its spatial-scale
may be ultimately limited by the phase-coherent length of the disordered
insulator. Evidence for quantum coherence in the Anderson-insulating phase may
be obtained from magneto-resistance measurements which however are useful
mainly in thin-films. The technique used in this work offers an empirical
method to measure this fundamental aspect of Anderson-insulators even in
relatively thick samples.

\end{abstract}
\maketitle

\section{Introduction}

Disorder may lead to a variety of non-trivial phenomena in both classical and
quantum systems. The most familiar of these phenomena is Anderson localization
\cite{1}. This phenomenon has been established in electronic transport
\cite{2}, propagation of light \cite{3} and sound waves \cite{4}, and in
disordered Bose-Einstein condensates \cite{5}.

Localization of wavefunctions may seem a way to allow manipulation of a
particular site in a solid while parts of the system that are remote from it
are unaffected. This expectation has been recently questioned; Khemani,
Nandkishore, and Sondhi (KNS) \cite{6} shown that adiabatically changing the
potential on a local site will produce an effect over a distance that may
exceed $\xi$ by a considerable margin. This long-range effect may have
important consequences for quantum-computing manipulations and for fundamental
issues such as the orthogonality-catastrophe \cite{6,7}.

In this work we describe a method that allows observation of the KNS effect in
an electronic system and show results that demonstrate the quantum nature of
the phenomenon.

\section{Experimental}

\subsection{Sample preparation}

The samples used in this study were amorphous indium-oxide (In$_{\text{x}}$O)
made by e-gun evaporation of 99.999\% pure In$_{\text{2}}$O$_{\text{3-x}}$
onto room-temperature Si-wafers in a partial pressure of 1.3x10$^{\text{-4}}%
$mBar of O$_{\text{2}}$ and a rate of 0.3$\pm$0.1\AA /s. Under these
conditions the carrier-concentration \textit{N} of the samples, measured by
the Hall-Effect at room-temperatures, was \textit{N}$\approx$(1$\pm
$0.1)x10$^{\text{19}}$cm$^{\text{-3}}$. Using free-electron formula, this
carrier-concentration is associated with $\partial$n/$\partial\mu\approx
$10$^{\text{32}}$erg$^{\text{-1}}$cm$^{\text{-3}}$. The Si wafers (boron-doped
with bulk resistivity $\rho\leq$2x10$^{\text{-3}}\Omega$cm) were employed as
the gate-electrode in the field-effect experiments. A thermally-grown
SiO$_{\text{2}}$ layer, 2$\mu$m thick, served as the spacer between the sample
and the conducting Si:B substrate. Films thickness was measured in-situ by a
quartz-crystal monitor calibrated against X-ray reflectometry. Samples
geometry was defined by the use of stainless-steel mask during deposition into
rectangular strips 0.8$\pm$0.1mm wide and 1$\pm$0.1mm long.

\subsection{Measurement techniques}

Conductivity of the samples was measured using a two-terminal ac technique
employing a 1211 ITHACO current preamplifier and a PAR 124A lock-in amplifier
using frequencies of 30-75Hz depending on the RC of the sample-gate structure.
R is the source-drain resistance and C is the capacitance between the sample
and the gate (C in our samples was typically $\cong$10$^{\text{{\small -10}}}%
$F and R for the samples studied in this work ranged between 1.5-20M$\Omega$).
Except when otherwise noted, the ac voltage bias in conductivity measurements
was small enough to ensure near-ohmic conditions. Most measurements were
performed with the samples immersed in liquid helium at T$\approx$4.1K held by
a 100 liters storage-dewar. This allowed up to two months measurements on a
given sample while keeping it cold (and in the dark). These conditions are
essential for measurements where extended times of relaxation processes are
required at a constant temperature. All samples described below were
Anderson-insulating and exhibited hopping conductivity $\sigma$ that for 4%
$<$%
T%
$<$%
50K was of the Mott form \cite{8};
\begin{equation}
\sigma(T)\approx\exp\left[  -\left(  \text{{\small T}}_{\text{0}%
}\text{{\small /T}}\right)  ^{\text{1/4}}\right]
\end{equation}
as illustrated in Fig.1 for two typical samples.
\begin{figure}[ptb]%
\centering
\includegraphics[
height=2.2658in,
width=3.039in
]%
{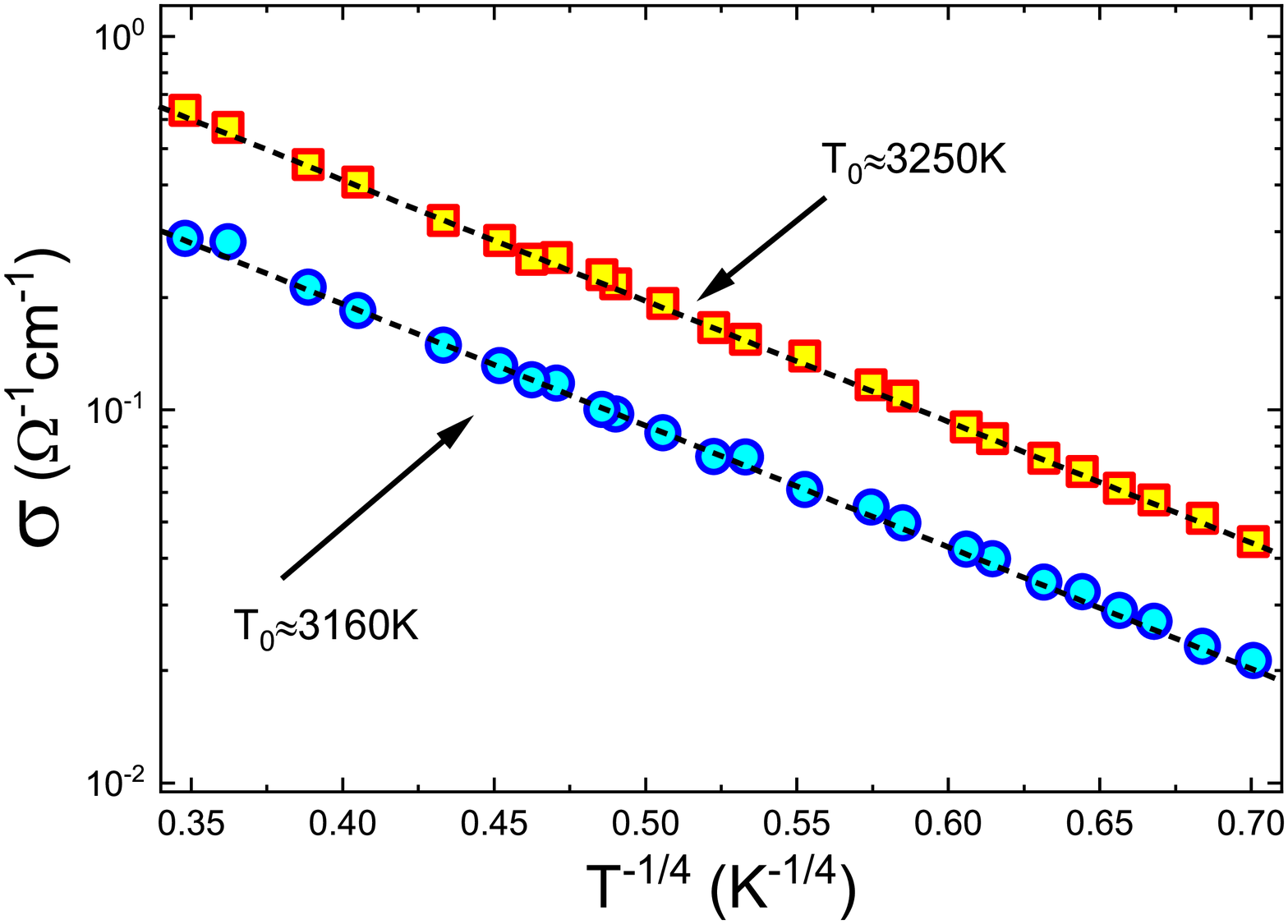}%
\caption{Conductivity versus temperature for two In$_{\text{x}}$O samples from
the same preparation batch with thickness d=82 nm. These exhibit Mott
variable-range-hopping yielding similar activation energies T$_{\text{0}}$ and
localization lengths $\xi\approx$3nm (see text for details).}%
\end{figure}
This allowed an estimate of the localization-length $\xi$ through \cite{8}:
k$_{\text{B}}$T$_{\text{0}}\approx$($\xi^{\text{3}}\partial$n/$\partial\mu
$)$^{\text{{\small -1}}}$~where $\partial$n/$\partial\mu$ is the thermodynamic
density-of-states. With $\partial$n/$\partial\mu\approx$10$^{\text{32}}%
$erg$^{\text{-1}}$cm$^{\text{-3}}$, the $\xi$ values for the samples reported
below ranged between 2.7nm to 3.3nm. These $\xi$ values are close to the
inter-carrier distance N$^{\text{-1/3}}$of this version of In$_{\text{x}}$O as
may be expected for samples that are far from the metal-insulating transition
which applies to all our studied samples. This makes the estimate for $\xi$,
based on the $\sigma$(T) data, a plausible value.

Taking the sample far from equilibrium to study its thermalization dynamics is
accomplished in this work by exposing the sample to an AlGaAs diode operating
at $\approx$0.82$\pm$0.05$\mu$m mounted on the sample-stage $\approx$10-15mm
from the sample. The diode was energized by a computer-controlled Keithley 220
current-source. Upon exposure to the infrared source, the electrons are
promptly raised to a high energy state and their excess energy is then
dissipated into the phonon system (a radiation-less process \cite{9}). In this
method, only the sample is efficiently heated and its excess energy is
uniformly distributed throughout the sample much faster than the time-scale of
the experiments described below. Full details of this technique and its
application for the study of several Anderson insulators are described in
\cite{9}.

\section{Results and discussion}

In a field-effect experiment, the charge $\delta$Q added to the sample when
the gate-voltage is changed by $\delta$V$_{\text{g}}$, resides in a thin layer
of thickness $\lambda\approx$(4$\pi$e$^{\text{2}}\partial$n/$\partial\mu
$)$^{\text{-1/2}}$ at the interface between the sample and the spacer
\cite{10,11} (The dielecric constant of the material $\kappa$ is of the order
of 10). The thickness of this layer in In$_{\text{x}}$O is $\lambda\approx
$2nm, which is much smaller than the thickness d of the samples studied in
this work that ranged between 45nm to 150nm. Yet, it turns out that the added
charge $\delta$Q to the system due to $\delta$V$_{\text{g}}$ had an effect
extending over length-scales much longer than both $\lambda~$and $\xi$.

This observation may be inferred from G(V$_{\text{g}}$) plots, taken at
different times t$_{Q}$, while the sample is relaxing after being
{\small quench-cooled} from an excited state; Consider the G(t,V$_{\text{g}}$)
plots for the samples in Fig.2. The protocol used throughout the series of
measurements shown in this composite figure was as follows. The sample,
immersed in liquid He at T=4.11K, was exposed for 3 seconds to infra-red
source (light-emitting-diode at 0.82 micron radiation) taking it from
equilibrium. G(V$_{\text{g}}$) scans were then taken with constant $\partial
$V$_{\text{g}}$/$\partial$t starting from V$_{\text{g}}$=0, at which the
gate-voltage was kept between subsequent scans. These are labeled in the
graphs by the time t$_{Q}$ that elapsed since turning off the infra-red source
(and the onset of relaxation towards restoring equilibrium under V$_{\text{g}%
}$=0). Each of these G(V$_{\text{g}}$) plots reflects the energy dependence of
$\partial$n/$\partial\mu$ modulated by a "memory-dip" which results from the
interplay between disorder and Coulomb interaction \cite{12}.%
\begin{figure}[ptb]%
\centering
\includegraphics[
height=5.0332in,
width=3.039in
]%
{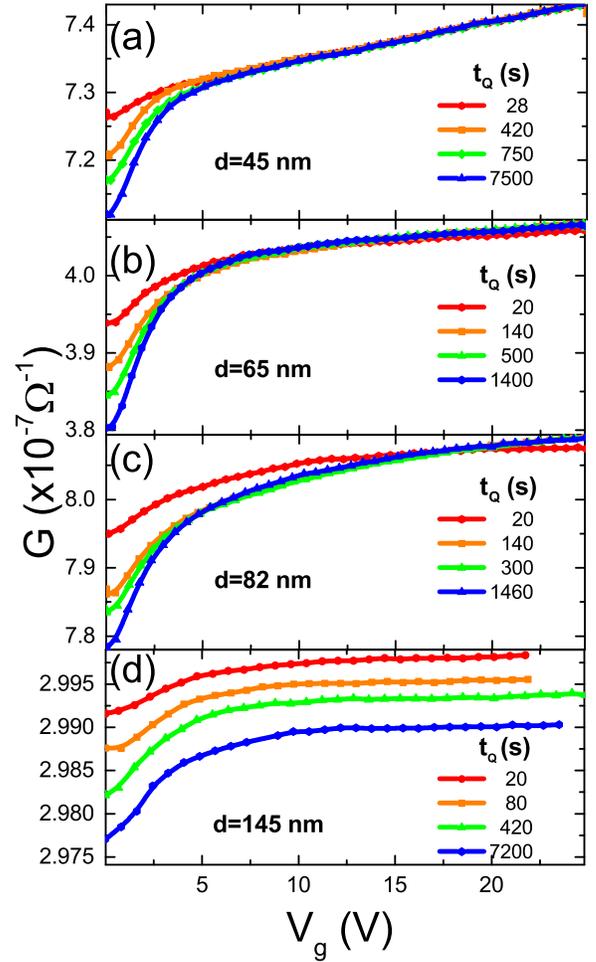}%
\caption{ Results of the G(t,V$_{\text{g}}$) protocol performed on
In$_{\text{x}}$O samples with similar composition (carrier-concentration of
$\approx$10$^{\text{19}}$cm$^{\text{-3}}$) but different thickness d. Each
G(V$_{\text{g}}$) plot were obtained with the same sweep-rate $\partial
$Vg/$\partial$t=0.5V/s. Bath temperature T=4.1K}%
\end{figure}

Note first the difference between the thinnest and thickest samples in the
series, Fig.2a and 2d respectively. In the former, the G(V$_{\text{g}}$) plots
taken at different times tend to merge for V$_{\text{g}}\geq$10V while, for
the 145 nm sample, they tend to become parallel.

A simple explanation to the results exhibited by the sample in Fig.2d is that
the added charge only affects the part of the sample that is close to the
spacer-interface while the rest of it is unaffected. In this case the sample
is effectively composed of two conductors in parallel; one where the
G(t,V$_{\text{g}}$) curves are like the pattern exhibited by the sample in
Fig.2a, and another for which G(t) just monotonically decreases after the
quantum-quench, independent of $\delta$V$_{\text{g}}$. Superimposing these two
components qualitatively reproduces the G(t,V$_{\text{g}}$) curves exhibited
by the 145 nm sample in Fig.2d.

It is important to understand the different roles played by the infrared
exposure versus the gate-sweeps in these experiments. Changing the
gate-voltage or exposing the sample to infrared will take the system from
equilibrium. However, these agents do not play a symmetric role in the
protocol; the infrared exposure is a one-shot event driving the system far
from equilibrium. Sweeping the gate is used to take a snapshot of how far the
system is on its relaxation trail. This is done intermittently as time
progresses and yields a certain swing $\delta$G(t) reflecting the development
of a memory dip. This $\delta$G may be then compared with the background
conductance-value that is going down with time due to the original excitation
by the infrared source (which, as alluded to in section III, affects the
entire thickness of the sample). The form of the observed G(t,V$_{\text{g}}$)
plots will tell whether or not the gate-sweeping affects the entire sample
volume (as in Fig.2a, 2b and 2c) or only part of it (as in Fig.2d).

The G(t,V$_{\text{g}}$) curves pattern characteristic of a thin sample has
been first observed in \cite{13} on crystalline indium-oxide and later in
\cite{14} on a different version of In$_{\text{x}}$O than the one used here
(namely, with \textit{N}$\approx$8x10$^{\text{19}}$cm$^{\text{-3}}$).

To account for the behavior of the three thinner samples is a more challenging
task; apparently in these instances the disturbance caused by the added charge
extends throughout their entire thickness - over a length-scale of d which,
for the 82nm sample, is 25 to 30 times larger than the localization-length
$\xi$. It is hard to see how such a long-range effect is possible unless
wavefunction-overlap that are L$\gg\xi$ apart is much better than might be
expected from exponential decay. The Coulomb interaction due to $\delta$Q over
this length-scale, even if unscreened, is too weak relative to the local
disorder to affect G(V$_{\text{g}}$) during the time V$_{\text{g}}$ is swept.

High transmission-channels through disordered media would offer an explanation
for the long-range effect. These resonant channels are theoretically possible
but exponentially rare \cite{15,16,17}. By contrast, the scenario proposed by
KNS creates such resonant channels in the disordered system with high
probability by using a time-dependent adiabatic process \cite{6}. Adapted for
our geometry, quasi-extended states are parametrically formed perpendicular to
the film plane by slowly varying the local potential V at the interface layer.
As will be now shown, this scenario accounts for all aspects of the
experimental results.

Let us first look at r$_{\text{zd}}$, the extent of the `zone-of-disturbance'
expected of the KNS-produced resonances \cite{6}:%
\begin{equation}
\text{r}_{\text{zd}}\approx\xi\text{\textperiodcentered}\ln\left(
\frac{\text{W}^{\text{2}}}{\partial\text{V/}\partial\text{t\textperiodcentered
}\hbar}\right)
\end{equation}
With the value of the quenched-disorder in our samples \cite{18} W$\approx
$0.5-1eV, rate of potential-change \cite{19} $\partial$V/$\partial$t$\approx
$0.5meV/s, and $\xi\approx$3nm Eq.1 gives r$_{\text{zd}}\approx$100nm.

Note that r$_{\text{zd}}\approx$100nm is consistent with our results (Fig.1);
it is close to the film thickness below which the G(t,V$_{\text{g}}$) curves
converge at high gate-voltages which implies r$_{\text{zd}}\geq$d. In
addition, the mean values that $\delta$V attains in the V$_{\text{g}}%
$-interval used in the experiments, covers the energy separation $\delta
$E$\approx(\partial$n$/\partial\mu$\textperiodcentered L$^{\text{3}%
})^{\text{-1}}$ for states that are apart by any L$\gtrsim$2$\xi$. This
secures ample `tuning-margin' for creating the quasi-extended states by the
KNS scenario.

A fundamental requirement on the KNS mechanism is that phase-coherence must be
preserved throughout the spatial-scale in question. This requirement follows
from the quantum-mechanical nature of the process. In other words, the range
of disturbance may be r$_{\text{zd}}$ in Eq.1 \textit{only} when L$_{\phi}$,
the phase-coherent-length in the medium obeys L$_{\phi}$%
$>$%
r$_{\text{zd}}$.

Evidence for phase-coherence in Anderson-localized films over scales of tens
of several nano-meters has been reported. This evidence is based on two
phenomena, both strictly requiring phase-coherence: orbital
magneto-conductance \cite{20,21}, and Andreev tunneling \cite{22}. The latter,
performed on In$_{\text{x}}$O films of similar composition as used in the
current work, demonstrated that a coherence-length of $\simeq$60nm at
T$\approx$4K is realizable in this system.

A further test of the role of quantum-coherence in the nonlocal effect
discussed here is to see how the G(t,V$_{\text{g}}$) plots change when
dephasing is judiciously introduced. Once the dephasing-rate is large enough
to cause L$_{\phi}$%
$<$%
d, the resulting G(t,V$_{\text{g}}$) plots should revert from the `converging'
pattern to that resembling the results in Fig.1d.

To implement this test in a controlled way, one needs a dephasing agent that
can be turned on and off at will. An effective and easy to control mechanism
for dephasing Anderson-insulators is using a non-ohmic field in the transport
measurement \cite{20}. This has been demonstrated in magneto-conductance
measurements on strongly-localized indium-oxide films \cite{20}. This
technique was applied on three different In$_{\text{x}}$O samples and the
results corroborate the expected behavior caused by the extra dephasing.
Figure 3 illustrates the results of one of these experiments:%

\begin{figure}[ptb]%
\centering
\includegraphics[
height=4.4659in,
width=3.039in
]%
{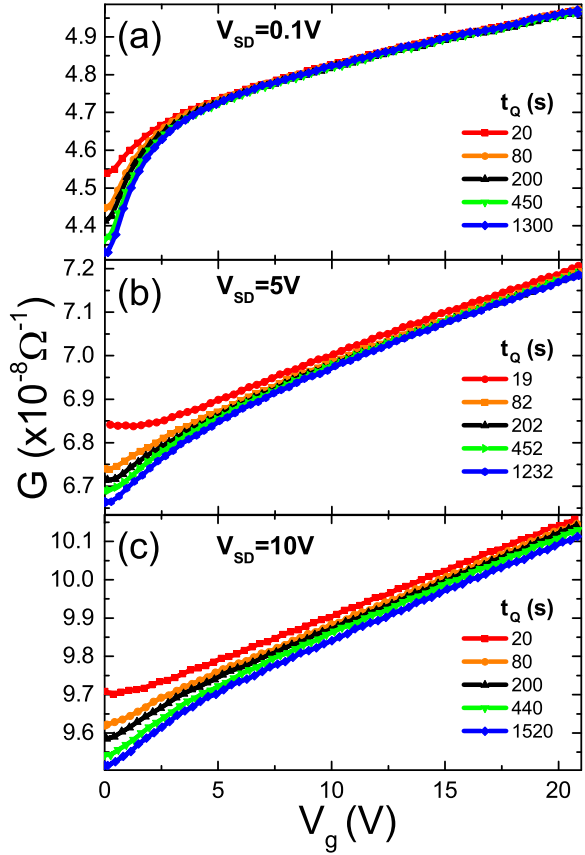}%
\caption{Results of the G(t,V$_{\text{g}}$) protocol applied on a single 65nm
thick In$_{\text{x}}$O sample under different source-drain fields (distance
between the source-drain contacts is 1mm). Plate (a) shows the results of the
protocol taken under linear-response conditions while in plates (b), and (c)
the G(t,V$_{\text{g}}$) plots were taken using non-ohmic voltages in the
measurement causing the somewhat enhanced conductance. Bath temperature
T=4.1K}%
\end{figure}

Figure 3a shows a set of G(t,V$_{\text{g}}$) curves taken in linear-response.
These `converging' plots are consistent with r$_{\text{zd}}$%
$>$%
d. Using non-ohmic V$_{\text{SD}}$ for measuring G(V$_{\text{g}}$) on the same
sample produced however, different results; the G(t,V$_{\text{g}}$) curves
(Fig.3b and 3c) resemble the pattern obtained for the thick sample in Fig.2d
where presumably the range of $\delta$V$_{\text{g}}$ is smaller than the
sample thickness.

Another indication that, under the higher V$_{\text{SD}}$ conditions, part of
the sample is not affected by the gate-voltage is shown in Fig.4.
\begin{figure}[ptb]%
\centering
\includegraphics[
height=2.1387in,
width=3.039in
]%
{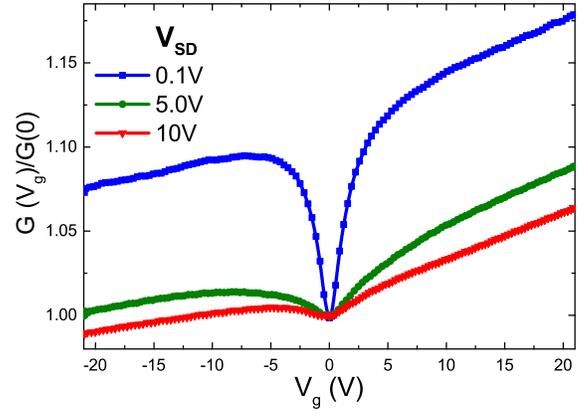}%
\caption{The memory-dips taken under the same source-drain voltages and
sweep-rates as the data in Fig.2. These were taken, in each case, after the
sample was allowed to relax at V$_{\text{g}}$=0V for 24 hours. The bath
temperature was T=4.1K.}%
\end{figure}
This figure compares the relative magnitude of the memory-dips taken under the
same fields used in Fig.3a, 3b, and 3c. The figure shows a large reduction in
the memory-dips magnitude for the two non-ohmic V$_{\text{SD}}$ used relative
to the linear response plot. The reduced range of disturbance implied by the
data in Fig.3b, 3c and 4 is consistent with the dephasing effect of non-ohmic
fields causing L$_{\phi}$ to become the shortest scale. Similar behavior was
observed on two other samples with d=65nm and d=82nm upon application of
non-ohmic fields.

For further discussion of the results of the non-ohmic fields, we show in
Fig.5 resistance versus source-drain voltage V$_{\text{SD}}$ plots for three
of the samples used in this study.%
\begin{figure}[ptb]%
\centering
\includegraphics[
height=2.1646in,
width=3.039in
]%
{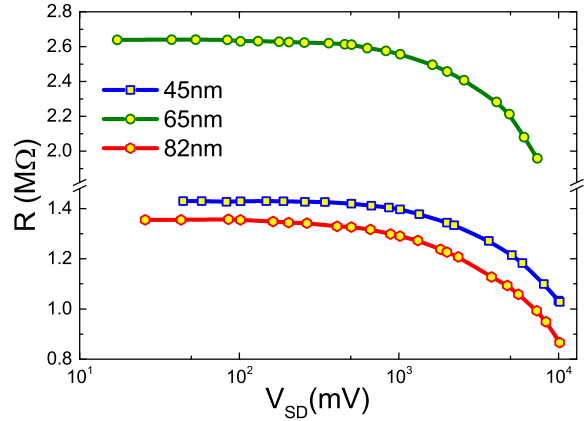}%
\caption{Sample resistance as function of the source-drain voltage
V$_{\text{SD}}$. Plots are given for three sample labeled by their thickness.
The nominal field used for ohmic-regime measurement was typically
F=1V\textperiodcentered m$^{\text{-1}}$. The source-drain separation for all
these samples was 1mm.The upper plot is taken on the same sample as in Fig.3
above.}%
\end{figure}

Note first that the increase in the overall conductance of the sample used in
Fig.3 under V$_{\text{SD}}$ of 5V and 10V (by a factor of $\approx$1.5 and
$\simeq$2 respectively, see Fig.5 ) is not the reason for the qualitative
change in the G(t,V$_{\text{g}}$) plots; when measured in linear-response,
samples with the same d, but conductance that differed by as much as order of
magnitude, still exhibited the same converging G(t,V$_{\text{g}}$) curves.
Secondly, in terms of dephasing the effect of a non-ohmic field act in the
same direction as higher sample temperature \cite{20}. The increase of
effective temperature $\Delta$T due to the applied source-drain field
F$_{\text{{\small SD}}}$ may be roughly estimated as $\Delta$T$\simeq$e$\xi
$F$_{\text{SD}}$/k$_{\text{B}}$which, for the F$\approx$%
10Vm$^{\text{{\small -1}}}$used in Fig.3c is tantamount to $\Delta$T$\approx$0.1-0.3K.

That L$_{\phi}$ in an Anderson-insulator may be $\geq$82nm at T$\approx$4K,
implicit to our proposed picture, is not an obvious fact, it deserves some
elaboration; To put things in perspective, L$_{\phi}$ of this order of
magnitude is typical of \textit{diffusive} samples at this temperature
\cite{23}. This may conflict with common intuition expecting disorder to
decrease transport-related spatial-scales. However the dependence of L$_{\phi
}$ on disorder is not clear even in diffusive systems despite extensive
studies \cite{24}, let alone in the more intricate Anderson-insulating phase
where this issue has barely been studied. On the basis of current knowledge it
is not impossible that L$_{\phi}$ in the insulating phase be as large as in
the metallic phase. In terms of mechanisms, the insulating phase may even have
an advantage; electron-electron inelastic scattering that, at low
temperatures, is the main source of dephasing in diffusive systems, is
suppressed in the insulating phase. This has been anticipated on theoretical
grounds \cite{24}, and was shown experimentally \cite{25}. Moreover, the
electron-phonon inelastic rate is likely also suppressed due to the reduced
overlap between the initial and final electronic states involved in the
inelastic event. Therefore dephasing due to inelastic scattering may actually
be weakened in the strongly-localized regime. On the other hand, once
interactions are turned on, a potential source for dephasing appears that may
not have existed in the weak-disorder regime - spin-flips. This mechanism may
become important once the on-site Coulomb repulsion is strong enough to
precipitate a finite density of singly-occupied states at the Fermi-energy
\cite{26}. These singly-occupied sites act like local magnetic impurities and
may contribute to dephasing \cite{27}. This potential source of dephasing may
be the reason for the paucity of experiments reporting on quantum-interference
effects in Anderson insulators in systems that do exhibit such effects in
their diffusive regime. Evidence for quantum-coherent effects is usually based
on observation of anisotropic magneto-conductance. This technique however
becomes ineffective for films thicker than few tens of a nanometer \cite{21},
a weakness not shared by our protocol.

In sum, we demonstrated the existence of a nonlocal effect in strongly
disordered Anderson-insulators extending over surprisingly long
spatial-scales. It was shown that this effect is consistent with the mechanism
proposed by KNS. The study also revealed that this spatial-scale is limited by
the phase-coherent length of the medium. Therefore the KNS effect is expected
to be considerably weakened by temperature while being only logarithmically
sensitive to the rate-dependence of a local potential change. Inasmuch as
dephasing is dominated by inelastic-scatterings, the phase-coherent length in
the insulating phase may be longer than intuitively expected. This is actually
a natural outcome of localization-induced discreteness. In the presence of
on-site interaction however, spin effects may become important and
phase-coherence could be compromised, even in the absence of inelastic events,
depending on the nature of the spin system \cite{27}. The non-equilibrium
technique employed in this study may offer a way to experimentally study these
fundamental issues of disordered quantum systems.

\begin{acknowledgments}
The author is grateful for the stimulating discussions with participants in
the "\textit{The Dynamics of Quantum Information}" program held in 2018 at the
Kavli Institute, Santa Barbara. This research has been supported by a grant No
1030/16 administered by the Israel Academy for Sciences and Humanities.
\end{acknowledgments}


\begin{thebibliography}{99}
\bibitem {19}For the SiO$_{\text{2}}$spacer-thickness $\delta$=2$\mu$m used in
these experiments and the thickness of the screeniong length $\lambda\approx$2
nm, the $\delta$V at the sample-interface layer is $\approx$10$^{\text{-3}}%
$x($\delta$V$_{\text{g}}$).
\end{thebibliography}
\end{document}